# CLIMATOLOGY OF THE MARTIAN POLAR REGIONS: THREE MARS YEARS OF CRISM/MARCI OBSERVATIONS OF ATMOSPHERIC CLOUDS AND DUST.

A.J. Brown[1] and M. J. Wolff[2] [1]SETI Institute, 189 N. Bernardo Ave Mountain View, CA 94043, abrown@seti.org, [2]Space Science Institute (18970 Cavendish Rd, Brookfield, WI, 53045). Author website: http://abrown.seti.org

**Introduction:** Here we continue the work started in {Brown, 2012;,Brown, 2] to document the dust and ice opacity of the atmosphere in the polar regions for Martian Years 28/29 and 30.

CRISM has been used to map the surface $CO_2$ and $H_2O$ ice cap springtime recessions for the north and south polar cap [3,4]. Grain size estimates were made of the surface ice, enabling us to constrain models of surface composition for the purposes of modeling the overlying atmosphere.

CRISM has the ability to take 'gimballed' observations of the surface as it passes over a target, thus creating what is termed an Emission Phase Function 'EPF' measurement (Figure 1) [5]. We report here on our initial investigations of the EPF polar observations and our attempts to model dust and ices suspended in the atmosphere and soil and ice covered surface.

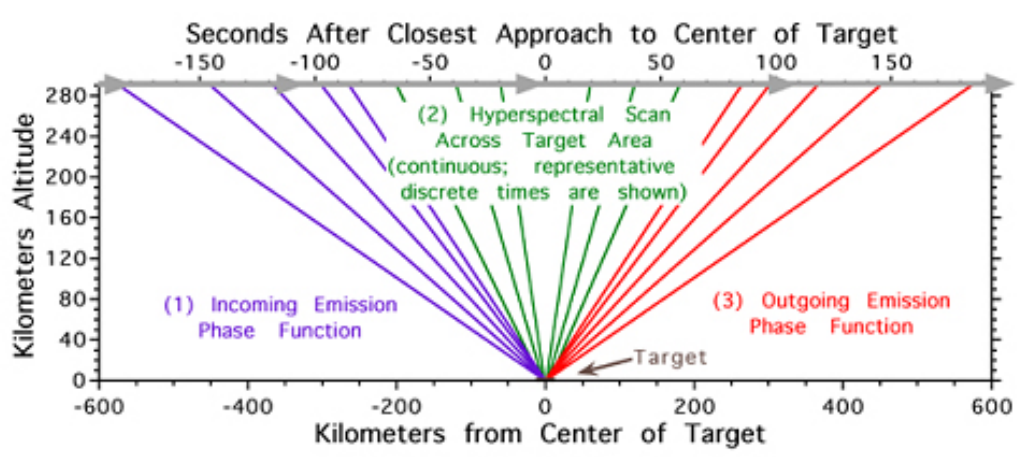

**Figure 1.** Schematic of a CRISM EPF observation.

| Earth DOY | MY/$L_s$ | *EPF* | *FRT* | *HRL* | *HRS* |
|---|---|---|---|---|---|
| 06_272-298 | 28/[113-125) | *21* | 6 | 5 | 1 |
| 298-312 | [125-132) | | | | |
| 312-326 | [132-139) | *10* | 8 | 2 | 13 |
| 326-340 | [139-146) | *6* | 7 | 1 | 1 |
| 340-354 | [146-153) | 15 | *13* | *4* | *7* |
| 354-07_003 | [153-161) | 20 | 3 | 2 | |
| 003-017 | [161-168) | *24* | *6* | 2 | 1 |
| 017-031 | [168-176) | 2 | 1 | 3 | 1 |
| 031-048 | [176-185) | 11 | 3 | 1 | |
| Northern winter | | | | | |
| --- | --- | *36* | 1 | 2 | 3 |
| 255-269 | [312-320) | *14* | 7 | 1 | |
| 269-283 | [320-328) | | 2 | *4* | 1 |
| 283-297 | [328-335) | | 1 | 2 | |
| 297-311 | [335-344) | | 1 | 1 | |
| 311-325 | [344-351) | | | | |
| 325-339 | [351-358) | | | | |
| 339-353 | [358-005) | | 9 | | |
| 353-08_002 | 29/[5-12) | | 38 | | |
| 08_002-016 | [12-19) | | 75 | 9 | 1 |
| 016-030 | [19-25) | | 122 | 10 | 3 |
| 030-044 | [25-32) | | *144* | 8 | |
| 044-058 | [32-38) | | 120 | 18 | 7 |
| 058-072 | [38-44) | | 159 | 18 | 38 |
| 072-086 | [44-50) | | 9 | 1 | |
| 086-100 | [50-56) | | | | |
| 100-114 | [56-62) | | 50 | 4 | *7* |
| 114-128 | [62-69) | | *74* | 38 | 16 |
| 128-142 | [69-75) | | 29 | *15* | 2 |
| 142-156 | [75-81) | | 13 | 1 | |
| 156-170 | [81-87) | | *44* | *12* | |
| 170-184 | [87-93) | | 53 | 17 | 2 |
| 184-198 | [93-100) | | 43 | 59 | 12 |
| 198-212 | [100-106) | | 59 | 19 | 4 |
| 212-226 | [106-112) | | 22 | *34* | 9 |
| 226-240 | [112-119) | | 46 | *25* | 25 |
| 240-254 | [119-125) | | 47 | 14 | 21 |
| 254-268 | [125-132) | | 22 | 12 | 22 |
| 268-282 | [132-139) | | *32* | 4 | 7 |
| 282-296 | [139-146) | | 32 | 24 | 14 |
| 296-310 | [146-153) | | 21 | 22 | 11 |
| 310-325 | [153-160) | | 39 | 4 | 17 |
| 356-09_012 | [177-190) | | 4 | | |
| Northern winter | | | | | |
| 193-233 | [301-325) | | 2 | 1 | 1 |
| MRO in safe mode | | | | | |
| 10_031-039 | 30/[45-49) | | 37 | 15 | |

| Earth DOY | MY/$L_s$ | *EPF* | *FRT* | *HRL* | *HRS* |
|---|---|---|---|---|---|
| 06_352-5 | 28/[152-159) | 8 | | | |
| 006-016 | [160-168) | 9 | | | |
| 016-033 | [168-176) | 3 | | 3 | 1 |
| 033-044 | [176-183) | 10 | 6 | 3 | 4 |
| 044-059 | [183-192) | 20 | 17 | 1 | 6 |
| 059-073 | [192-200) | 8 | 20 | 1 | 5 |
| 073-086 | [200-208) | 2 | 14 | | 1 |
| 086-101 | [208-217) | 25 | 27 | | 24 |
| 101-115 | [217-225) | *17* | *29* | 11 | 12 |
| 115-116 | [225-234) | | 4 | 1 | |
| 132-142 | [234-243) | | 36 | | 3 |
| 142-156 | [243-252) | 27 | *42* | *5* | 8 |
| 156-171 | [252-261) | 3 | 37 | 7 | 2 |
| 171-185 | [261-270) | *12* | 48 | 4 | 1 |
| 185-198 | [270-278) | 124 | 30 | 2 | 2 |
| 199-212 | [278-286) | *43* | 30 | | |
| 213-225 | [286-295) | *49* | *137* | 1 | 6 |
| 227-240 | [295-303) | 22 | *111* | 2 | 2 |
| 241-255 | [303-312) | | *111* | 4 | |
| 255-269 | [312-320) | *16* | 66 | 3 | 4 |
| 269-283 | [320-328) | | *34* | 16 | 4 |
| 283-297 | [328-335) | | *49* | 5 | 13 |
| 297-311 | [335-344) | | 18 | 8 | 4 |
| 311-348 | [344-002) | | | | |
| 348-00 | 29/[002-012) | | 61 | 1 | 2 |
| 08_004-33 | [012-026) | | 43 | 3 | |
| 033-074 | [026-044) | | 39 | 3 | 2 |
| Southern winter | | | | | |
| 356-09_001 | [177-184) | | 8 | 2 | |
| 019-033 | [194-202) | | 14 | 1 | |
| 034-050 | [203-211) | | 13 | 8 | 16 |
| 051-067 | [214-223) | | 8 | 4 | 2 |
| 069-082 | [223-233) | | 46 | 17 | 10 |
| 083-097 | [233-242) | | 31 | 18 | 10 |
| 098-112 | [242-252) | | 28 | 30 | 4 |
| 113-126 | [252-261) | | 16 | 14 | 9 |
| 128-142 | [261-271) | | 37 | 8 | 10 |
| 143-154 | [271-278) | | 14 | 16 | 5 |
| 162-175 | [282-292) | | 15 | 2 | |
| 176-192 | [292-302) | | 16 | 14 | 12 |
| 193-206 | [302-310) | | 28 | 14 | 17 |
| 207-217 | [310-317) | | 15 | 20 | 5 |
| 223-237 | [319-327) | | 41 | | |
| Southern winter followed by MRO in safe mode | | | | | |
| **TOTAL** | ***n=2177*** | **399** | **1319** | **252** | **207** |

**Table 1.** Totals of CRISM observations relevant to this study. North polar observations are on the left, and south polar on the right. Counts in italics indicate some missing geometries. Each line corresponds to the two week MRO planning cycle. DOY column gaps are when CRISM collected no data at the south pole.

Table 1 shows all the CRISM EPF observations poleward of 55°. CRISM is limited to daytime observations and MRO is in a ~250km circular orbit that crosses the equator south to north at 1500 local Mars standard time.

**Retrieval Method:** We are using the DISORT [6] algorithm to simulate the interaction of the Martian atmosphere and surface with the incident solar radiation with the goal of extracting three parameters: surface Lambert albedo, dust and ice aerosol optical depth.At present, the retrieval is accomplished by fitting a CRISM EPF (in a least-squares sense) at a single wavelength (0.696 micron) – we will be moving to multiple wavelengths in the near future to more effectively discriminate between the dust and water ice atmospheric components.

Computational efficiency is obtained by using the Look-up table approach, where we have pre-computed (millions of) radiative transfer models that span the possible range of optical depth and aerosol values. Additional numerical details include the use of 16 streams for the discrete ordinates solution and aerosol phase functions are taken from [7,8].

Dust optical constants are from [7] and water ice optical constants were from [9].

**Results:** The $\tau_d$ results for CRISM EPF and the corresponding MARCI $\tau_{ice}$ retrievals from Mars Year 28/29/30 (2006-2011) are shown in Figure 2-3.

**Dust opacity:** We reported in [1] that the results for this period are consistent with background dust opacities of $\tau_d$=0.3-0.5 for the south polar region, with average excursions to 1.4 during the MY28 dust event. An updated retrieval model suggests that our estimates were low, although comparing Figure 3 with Figure 2 of [1] shows that the corrected peaks are scaled up in opacity (by a factor of 2), but identical in shape.

**Dust Interpole Interannual Comparison:** The CRISM EPF dataset south polar dust activity is far more pronounced and repeatable than the north polar data. We are still investigating potential bias in our dataset, but first glance the south pole dust activity is more sharply concentrated each year around $L_s$=270 (southern vernal equinox). The north pole has a similar peak at $L_s$=90 (northern vernal equinox). MY28 was considerably more dusty than MY29 and 30 in both poles.

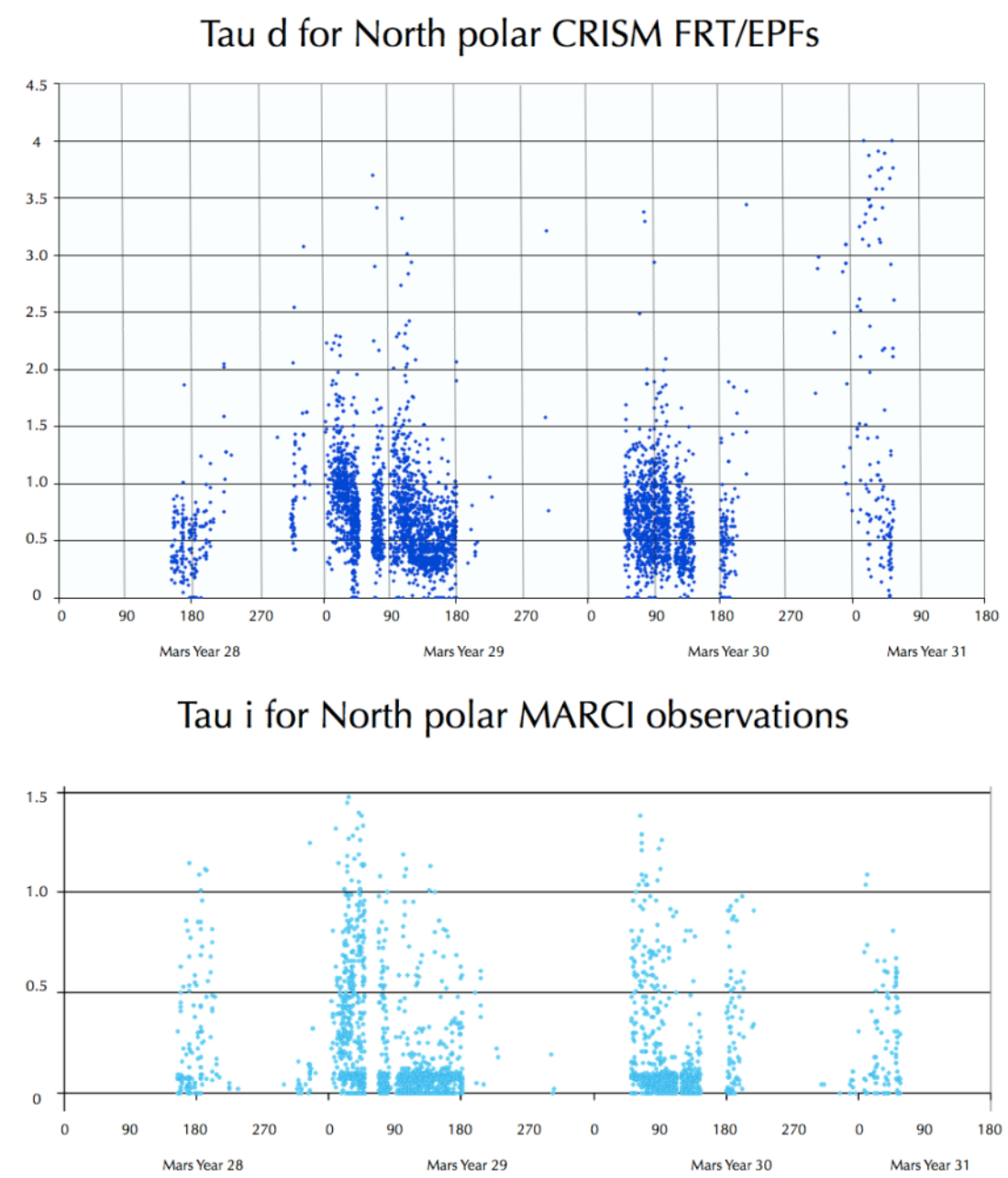

**Figure 2.** North polar $\tau_d$ and $\tau_i$ estimates for CRISM EPF/MARCI observations (poleward of 55°N) for MY28-30.

**Water ice opacity:** It should be noted that the MARCI retrievals presented here are only those corresponding to the CRISM EPF data – more complete MARCI analysis will be available at the time of the conference. Nervertheless, there does appear to be a peak in water ice opacity in due to the polar hood that reaches similar opacities (our model suggests peaks of $\tau_{ice}$ =1.5) in each pole that diminishes gradually during springtime. Background water ice is much lower than background dust opacity for both poles (consistently in 'background' measurements, $\tau_{ice} < 0.1 < \tau_d$).

**Conclusions:** We have presented estimates of dust opacity from CRISM and ice cloud opacity from MARCI in the north (Figure 2) and south pole (Figure 3) for the first three Mars years of MRO operations (MY28-30), which included a large dust event at MY28/$L_s$=260-270 and a smaller dust event in MY29 at the same time period. Future work will involve further investigations into MARCI water ice opacities throughout the year, filling in the gaps in our current observations.

**Acknowledgements:** Our thanks to Scott Murchie (CRISM PI) and the CRISM science operations team at JHU APL for their dedication to the task of obtaining this unique dataset. Our thanks to Mike Smith for providing his coork files for Martian $CO_2$ profiles.

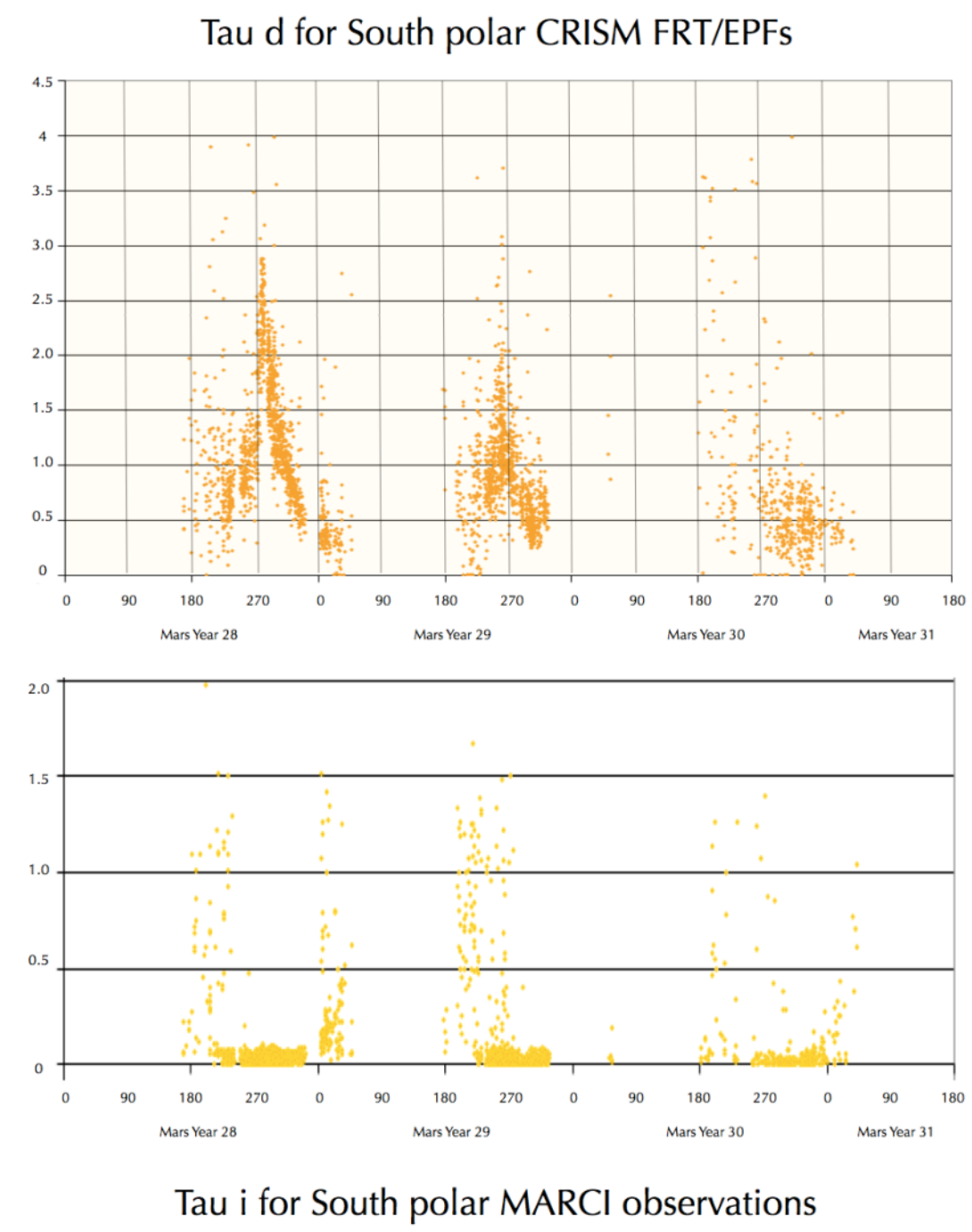

**Figure 3.** South polar $\tau_d$ and $\tau_i$ estimates for CRISM EPF/MARCI observations (poleward of 55°S) for MY28-30.

**References**

[1] Brown, A.J., Wolff, M.J. Atmospheric modeling of the Martian polar regions: One Mars year of CRISM EPF observations of the south pole. Lunar and Planetary Institute Science Conference Abstracts **40**, 1675 (2009).

[2] Brown, A.J., McGuire, P. Wolff, M.J. Atmospheric modeling of the martian polar regions: CRISM EPF coverage during the south polar spring recession. Lunar and Planetary Institute Science Conference Abstracts **39**, 2140 (2008).

[3] Brown, A.J., Calvin, W., McGuire, P. and Murchie, S. (2010) Compact Reconnaissance Imaging Spectrometer for Mars (CRISM) south polar mapping: First Mars year of observations. Journal of Geophysical Research. **115**, E00D13, doi:10.1029/2009JE003333 (2010).

[4] Brown, A.J., Calvin, W. and Murchie, S. Compact Reconnaissance Imaging Spectrometer for Mars (CRISM) north polar springtime recession mapping: First 3 Mars years of observations, Journal of Geophysical Research. **117**, E00J20, doi:10.1029/2012JE004113 (2012).

[5] Murchie S, Arvidson R, Bedini P, Beisser K, Bibring J-P, Bishop J, et al. Compact Reconnaissance Imaging Spectrometer for Mars (CRISM) on Mars Reconnaissance Orbiter (MRO). Journal of Geophysical Research. **112**, E05S3, doi:10.1029/2006JE002682 (2007).

[6] Stamnes K, Tsay SC, Wiscombe WJ, Jayaweera K. Numerically stable algorithm for discete-ordinate-method radiative transfer in multiple scattering and emitting layered media. Applied Optics. **27**, 2502-2509 (1988).

[7] Wolff MJ, Smith MD, Clancy RT, Arvidson RE, Kahre M, Seelos F, et al. Wavelength Dependence of Dust Aerosol Single Scattering Albedo As Observed by CRISM. Journal of Geophysical Research. **114**, doi:10.1029/2009JE003350 (2009).

[8] Wolff MJ, Clancy R.T., Cantor B.A. and Madeline, J.B. Mapping Water Ice Clouds (and Ozone) with MRO/MARCI. 4th International Workshop on the Mars Atmosphere. Paris, France. p. 213-6 (2011).

[9] Warren SG. Optical constants of ice from the ultraviolet to the microwave. Applied Optics. **23**,1206-25 (1984).